\documentstyle[preprint,prb,aps]{revtex}
\include{epsf}
\tighten

\begin{document}
\bibliographystyle{prsty}

\title{A Biased Monte Carlo Scheme for Zeolite Structure Solution}
\author{Marco Falcioni and Michael W.\ Deem}
\address{Chemical Engineering Department\\
University of California\\
Los Angeles, CA\ \ 90095-1592}
\maketitle
\begin{abstract}
We describe a new, biased Monte Carlo scheme to determine the crystal
structures of zeolites from powder diffraction data.  We test the
method on all publicly known zeolite materials, with success in all
cases.  We show that the method of parallel tempering is a powerful
supplement to the biased Monte Carlo.
\end{abstract}
\narrowtext
\section{Introduction}
Zeolites continue to be synthesized at a furious pace.  Crucial to the
development of the field of zeolite science is the ability to
determine the structure of newly-synthesized materials:
Structure is sought after not only to understand the performance of
newly synthesized catalysts but also to propose rational syntheses of
homologous materials with tailored performance.  Roughly 118 framework
structures have been reported, yet another several dozen distinct
synthetic zeolites remain unsolved in the patent literature.  Perhaps
soon the techniques of diversity synthesis will be introduced to the
field, with a tremendous explosion in the number of new, unsolved
synthetic zeolites.

\begin{figure}[p]
\epsfxsize=5in \centerline{\epsfbox{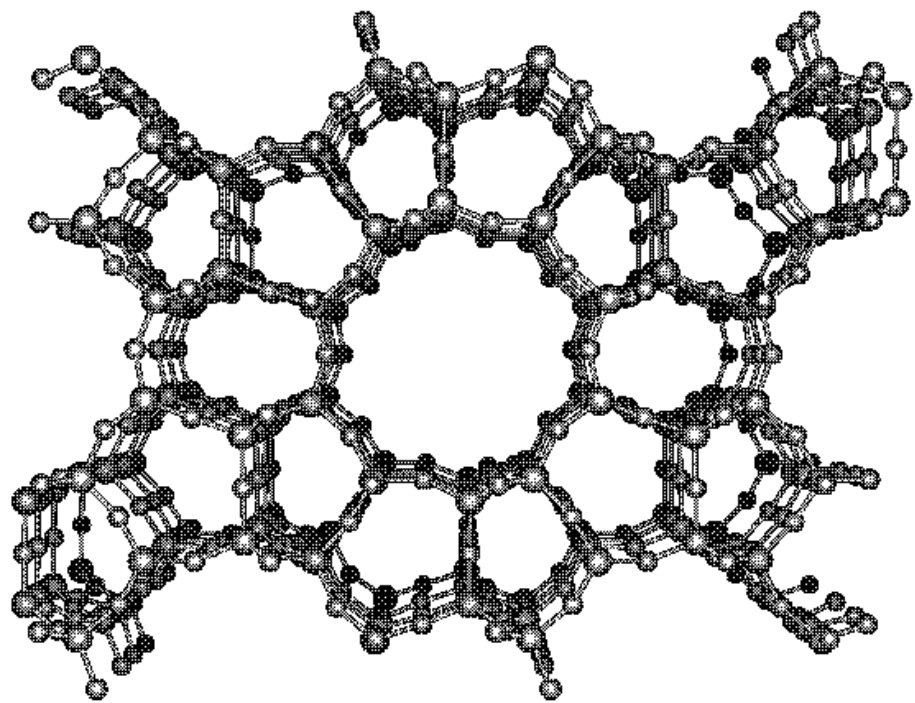}}
\caption{The framework structure of ZSM-5 (MFI), from Ref.\
\protect\onlinecite{Cerius2}.}
\label{fig:mfi}
\end{figure}

Zeolites are crystalline microporous materials that have found a wide
range of uses in industrial applications.  They are used as catalysts,
molecular sieves, and
ion-exchangers.\cite{III_Breck,III_Barrer,III_Ruthven,III_Rabo,III_Karge}
A typical example is ZSM-5, shown in Fig.~\ref{fig:mfi}, and used as a
cracking co-catalyst in the refinement of crude oil.  The pore
structure of this particular zeolite is 3-dimensional, as is that of
the zeolites chabazite, A, X, Y, and beta,\cite{atlas_book,I_Newsam}
but there are zeolites with 1-dimensional pores, such as cancrinite,
zeolite L, and AlPO$_4$-5, and 2-dimensional pores, such as
deca-dodecasil 3R, TMA-E(AB), and heulandite.  Classical zeolites are
aluminosilicates. The basic building block is a TO$_4$
tetrahedron. Usually T = Si, although substitution of the silicon with
aluminum, phosphorus, or other metals is common.  The tetrahedral
species is commonly denoted by T when one is concerned with
structural, rather than chemical, properties of the zeolite.

\begin{figure}[p]
\epsfxsize=5in \centerline{ \epsfbox{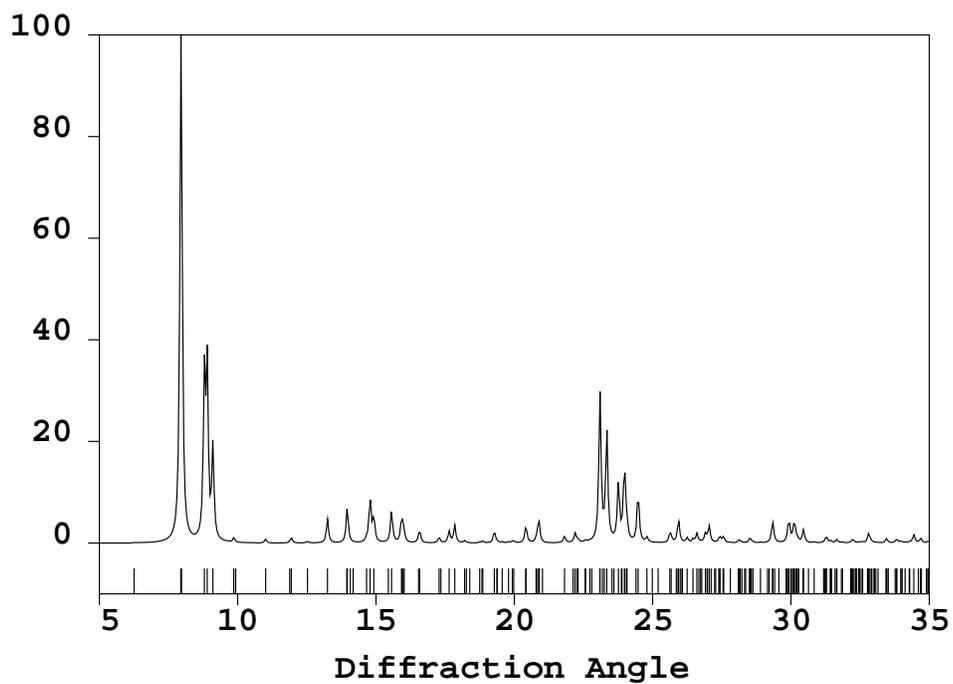}}
\caption{A simulated powder X-ray diffraction pattern for
ZSM-5 (MFI) framework, from Ref.\ \protect\onlinecite{Cerius2}.}
\label{fig:pxd}
\end{figure}

The derivation of an atomic-scale model of the framework crystal
structure of a newly-synthesized zeolite is a non-trivial task.  The
difficulty stems primarily from the polycrystalline nature of most
zeolite samples, with crystallite sizes typically below 5 $\mu$m.
Notable improvements in single-crystal diffraction techniques have
been made, primarily through the use of synchrotron X-rays, but
fundamental limitations still exist.\cite{III_NewsamVI,III_Mccusker} A
typical powder pattern is shown in Fig.~\ref{fig:pxd}.  Most zeolites
have been solved to date through physical model building efforts.  A
limited number of zeolites have been solved with conventional
crystallographic methods ({\em e.g.}, see Ref.~\onlinecite{Rudolf} for
a complex example).  These methods attempt to deconvolute the powder
diffraction data into unique reflections and then apply direct methods
to determine a structure.  This approach requires a high quality
scattering dataset for success, a dataset that usually is not
obtainable from zeolite samples.  Very few zeolite structures have
been solved via conventional crystallographic methods, which indicates
the limited applicability of this approach.  Recently, a few zeolites
have been solved with electron diffraction
methods,\cite{newsam88,davis97} but this is a laborious approach.

A direct, real-space method for zeolite structure solution from powder
diffraction data has been proposed.\cite{Deem89,Deem92a} The advantage
of this method is that it requires only easily available powder data,
and it incorporates little preconceived bias of the investigator in
the structure solution scheme.  In this approach, simulated annealing
with a simple Metropolis perturbation step is used to minimize a
zeolite figure of merit.  This approach has a nearly 90\% success rate
on materials with 6 or fewer crystallographically-distinct T-atoms.
To date, at least four other groups have used this approach to solve
new zeolite
structures.\cite{Akporiaye96a,Akporiayeb,Akporiaye96c,Stucky93,Cheetham,Vaughan}
On more complex structures, however, the method becomes unwieldy,
generating many hypothetical structures and often failing to find the
correct one.

In this paper, we apply powerful new ideas from Monte Carlo to the
problem of zeolite structure solution via powder diffraction data.
The key step is to define a cost function that is a function of the
atomic positions within the crystalline unit cell and that is
minimized by the structure corresponding to the experimental material.
Structure solution from powder data is a fundamentally challenging
problem, due to the presence of many local minima and large barriers
in the cost function.  We show that a combination of simulated
annealing and biased Monte Carlo is able to overcome the barriers in
most cases.  In the most difficult cases, we show that the new method
of parallel tempering is superior to simulated annealing and is able
to overcome the barriers.  This paper is organized as follows.  In
Section~\ref{sec:fom} we introduce the figure of merit for an unknown
zeolite sample. In Section~\ref{sec:montecarlo} we discuss the Monte
Carlo sampling of this figure of merit.  In Section~\ref{sec:results}
we show the results of applying our method to all publicly known
zeolites.  In Section~\ref{sec:discuss} we discuss the results of the
method and possible extensions.  We draw our conclusions in
Section~\ref{sec:conclusion}.

\section{The Figure of Merit}
\label{sec:fom}
For a given zeolite sample with known unit cell size, cell parameters,
symmetry, and density we want to construct a figure of merit. By
definition, the global minimum of the figure of merit should
correspond to the structure of the zeolite sample that we are
investigating.  The $n_{\rm unique}$ unique atoms are placed in the
unit cell.  Each of the $n_{\rm symm}$ symmetry operators generates an
atom position when applied to a unique T-atom, and so there are at
most $ n_{\rm unique} \times n_{\rm symm}$ atom positions in the unit
cell.  The figure of merit is a function of all of these parameters.
We keep fixed the cell parameters, the space group symmetry, and the
number of unique T-atoms: the only variables are the positions of the
T-atoms in the unit cell.  The figure of merit is defined as
\begin{eqnarray}
H & = & \alpha_{\rm T-T} H_{\rm T-T} + 
\alpha_{\rm T-T-T} H_{\rm T-T-T} \nonumber \\
& + & \alpha_{\langle {\rm T-T-T}\rangle} H_{\langle {\rm T-T-T}
\rangle} +  \alpha_{\rm D} H_{\rm D} + \alpha_{\rm M} H_{\rm M} +
\alpha_{\rm L} H_{\rm L} \nonumber \\ 
& + &\alpha_{\rm NB} H_{\rm NB} + \alpha_{\rm PXD} H_{\rm PXD} + 
\alpha_{\rm PND} H_{\rm PND} \ .
\label{eq:figure}
\end{eqnarray}
The figure of merit is defined for any arrangement of T-atoms, even
ones that are far from resembling a zeolite. The lower the value of
$H$, the more the structure resembles a zeolite of the given cell size,
symmetry, and density.  In general, one starts from a random configuration of
T-atoms and seeks the minimum of $H$ by moving the T-atoms suitably.
Each term in $H$ represents a particular contribution, and the
$\alpha_i$ are the relative weights. These weights are optimized
according to the rate of success of the method on one trial structure
and are generally kept fixed in the following.

It must be stressed that $H$ is not the thermodynamic energy.  For
one, we neglect the bridging oxygens and any cations or adsorbed
molecules that may be present in the real zeolite.  Furthermore, we
will see that the geometric terms of $H$ do not describe interactions
or pseudo-interactions, but are constructed to reproduce a
distribution of distances and angles observed in known zeolites. We
enforce the space group symmetry, and so symmetry-related T-atoms move
collectively.  In addition, we enforce crystalline order so that we
can limit our description of the material to a single unit cell.

One may wonder whether the explicit inclusion of bridging oxygens, or
whether a more detailed, quantum mechanical description of the system,
may be needed for an accurate description of the zeolite.
In fact, Eq.~(\ref{eq:figure}) captures the
relevant features necessary to describe and to predict the framework
structure of the zeolite crystal.  Once the positions of the T-atoms
are available, one can use more refined methods and more detailed
models to refine the structure or to study other properties that may
be of interest.

\begin{figure}[p]
\epsfxsize=2in \centerline{ \epsfbox{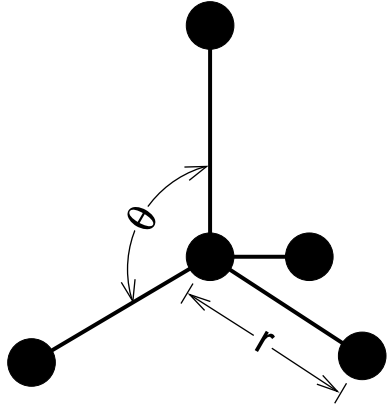}}
\caption{A typical tetrahedral structure: $\theta$ is the T--T--T
angle, and $r$ is the T--T distance.}
\label{fig:tetra}
\end{figure}

\begin{figure}[p]
\epsfxsize=5in \centerline{ \epsfbox{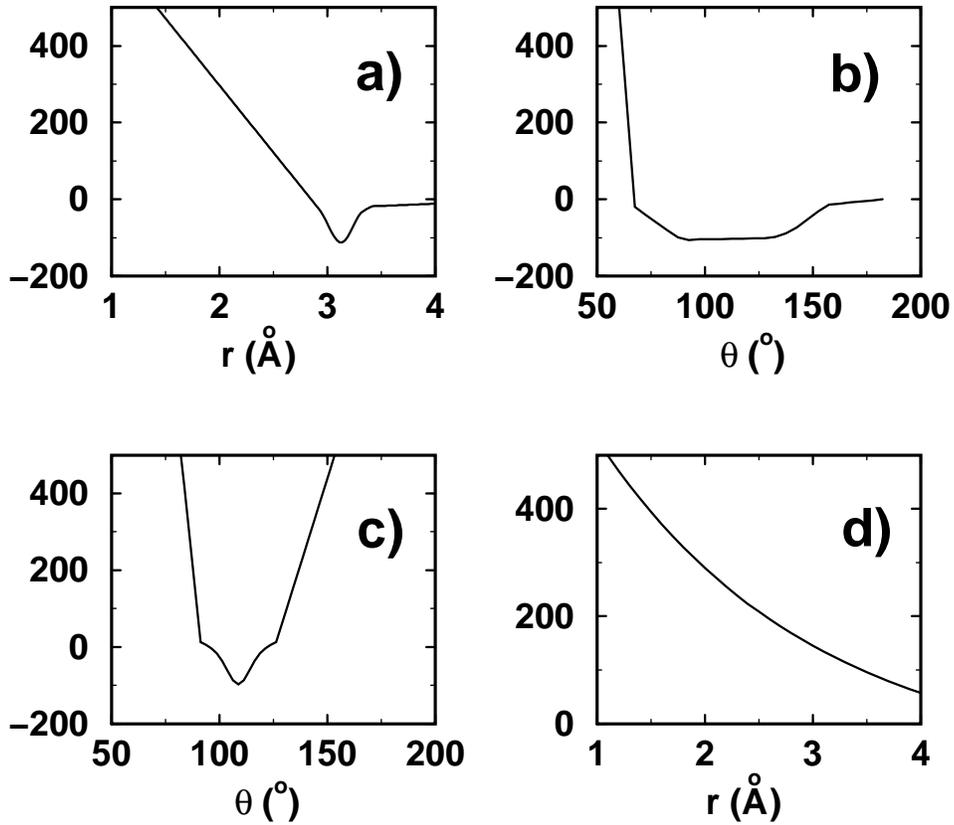}}
\caption{a) The T--T potential energy, b) the T--T--T angle potential
energy, c) the $\langle${T--T--T}$\rangle$ average angle potential
energy, and d) the repulsive potential energy for non-bonded
neighbors.}
\label{fig:pot}
\end{figure}

The different contributions to Eq.~(\ref{eq:figure}) are of three
types: the geometric terms, the density terms, and the diffraction
terms.  The first three terms of $H$ are geometric, and they are
obtained by histogramming the T--T distances and the T--T--T angles of
32 known high silica zeolites.\cite{III_NewsamI,Deem92a}  The T--T
distance is sharply distributed around 3.1 \AA, and the T--T--T angles
are distributed around 109.5$^{\rm o}$, as one would expect for
a tetrahedrally coordinated species. The angles and distances are shown in
Fig.~\ref{fig:tetra}.  The angle $\langle$T--T--T$\rangle$ is
the average of all of the angles around a given T-atom.  We are interested in
sampling configuration space according to Boltzmann
statistics.  This leads us to define potential energies that reproduce
the observed histograms when sampled according to $\exp(-\beta H)$
at a particular value of the inverse temperature $\beta = 1/T$. For
simplicity, we set Boltzmann's constant to unity. In
practice we take the logarithm of the histogrammed data and fit a
spline through them.  The potential energies are then extended to
ranges of angles and distances beyond the observed ones. The
resulting curves are shown in Fig.~\ref{fig:pot}a, \ref{fig:pot}b, and
\ref{fig:pot}c.  The correct histograms are obtained only when the
distribution is sampled at the inverse temperature $\beta$ used for the
inversion.  This value is arbitrary and simply defines what low
temperature means.

The $H_{\rm NB}$ and $H_{\rm L}$ terms account for the 4-connectedness
of silicates.  These terms are defined to be non-zero and positive
whenever a T-atom happens to have greater than or fewer than 4 first
neighbors respectively, where a neighbor is defined as an atom closer
than 4~\AA. If the number of neighbors is fewer than 4, we simply
assign a progressively larger weight to the atom.  Following
Ref.~\onlinecite{Deem92a} we use the values in Table~\ref{tbl:bond}.
The case in which there are more than 4~neighbors is treated
differently.  We define a list of the $N$ neighboring atoms.  We
choose 4~of them and treat them as connected neighbors, adding the
contributions from the potential energies discussed above.  We include
a repulsive potential energy, Fig.~\ref{fig:pot}d, for the remaining
unconnected neighbors. Since it is important to choose the best
connected neighbors for a configuration, we exhaustively search for
the combination of 4~connected and $N-4$ unconnected atoms that
minimizes the energy associated with the central $T$ atom.  In our
experience, this procedure gives better results than does simply
picking the neighbors according to their distance or bonding all
neighbors indiscriminately and assigning an extra weight to
over-coordinated atoms.

The $H_{\rm M}$ term favors merging.  Merging occurs in crystals
whenever a particular atom sits on a special position, a position
invariant under one or more symmetry operations other than the identity.
Since our method assigns positions in a stochastic way, it is unlikely
that we would find an atom exactly on a special position.  Therefore,
we define a merging range with a typical value of $r_{\rm M} = 0.8$~\AA.
Two or more symmetry-related atoms that fall within this distance are
merged.  When this condition occurs, all the atoms that fall within
the range are replaced by a single atom at the position of their
center of mass. The merged position is not a new position
for the system; it is merely the position used for the calculation of
the figure of merit.  $H_{\rm M}$ gives a negative, favorable
energy to merged atoms. This energy is linearly proportional to the merging
distance, {\em i.e.}\ the distance between the original and the
merged position.  Merging must be favored since the figure of merit is
proportional to the number of atoms in the unit cell, and placing an
atom on a special position lowers the number of actual atoms in the
unit cell and will typically remove a favorable energy contribution.
Merging is necessary whenever the number of T-atoms derived from the
experimental density, $n_{\rm o}$, is fewer than the number created
by symmetry, $n_{\rm unique} \times n_{\rm symm}$.  Merging is
disallowed whenever $n_{\rm o} = n_{\rm unique} \times
n_{\rm symm}$.  If merging is allowed, the number of T-atoms in the
unit cell can change.  In fact since the contribution $H_{\rm M}$
that favors merging is negative, there are situations in which some
atoms collapse on a highly symmetric special position, lowering the
density far below the observed one.  In order to enforce the observed
density we include the term
\begin{equation}
H_{\rm D} = (n_{\rm T} - n_{\rm o})^2 \ ,
\label{eq:density}
\end{equation}
where $n_{\rm T}$ is the actual number of atoms in the unit cell
after merging.

We now discuss some technical details of the figure of merit thus
defined.  Symmetry can be exploited in the computation of the figure
of merit in several ways.  Symmetry-related atoms will have the same
geometric structure surrounding them, so the energy is simply $n_{\rm
symm}$ times the energy of a single atom, unless merging has occurred.
In any case, the energy is proportional to the actual number of atoms
in the unit cell.  To speed up the calculation of the energy, we
divide the unit cell with a grid.\cite{sedgewick88} Each atom belongs
to a box in the grid, and the grid is designed so that first neighbors
to an atom will be in the same box as the atom or in one of the 26
neighboring boxes.  The computational effort to calculate the energy
is reduced from $O(n_{\rm unique}^2)$ to $O(n_{\rm unique})$ through
use of the grid.  The grid must be updated every time an atom is
moved, but since only one unique atom is moved at one time, we create
$n_{\rm unique}$ distinct grids.  This allows us to update only the
grid associated with a moved atom, leaving the others unchanged and
further reducing the computational effort used per attempted move.
The figure of merit we defined is highly non-local.  Symmetry
operations applied to the position of the unique atom can, in
principle, generate atoms anywhere in the unit cell.  This means that
the energy for all of the unique atoms must be recalculated every time
a single unique atom is moved.

The diffraction terms, $H_{\rm PXD}$ and $H_{\rm PND}$, incorporate in
the figure of merit experimental information that may be available
about the zeolite.  A typical PXD pattern is shown in
Fig.~\ref{fig:pxd}.  Let us assume that a skilled crystallographer has
collected high resolution powder X-ray scattering data on a zeolite
powder sample and has succeeded in indexing the resulting pattern.  We
then have available a list of Bragg reflections with Miller indices
$(hkl)$ and relative intensities.  For a given arrangement of atoms in
our model unit cell, we can compute the relative intensities for the
same list of reflections using standard formulas.\cite{Warren} The
intensity of a reflection, in arbitrary units, is given by
\begin{equation}
I(hkl) = p(\theta) \vert F_{hkl} \vert^2 \ ,
\label{eq:intensity}
\end{equation}
where $2\theta$ is the angle of the Bragg reflection, and the first
factor is the polarization term. The polarization term is $p(\theta) =
[1 + \cos^2(2\theta)]/[2\sin(\theta)\sin(2\theta)]$ for X-rays and
$p(\theta) = 1/[2\sin(\theta)\sin(2\theta)]$ for neutrons.  The
scattering amplitude is\cite{Warren}
\begin{equation}
F_{hkl} = \sum_{j=1}^{n_{\rm T}} f_j({\bf k}) \, o_j \,
\exp(-B_j\, {\bf k}^2 /4) \, \exp(i\, {\bf k} \cdot {\bf x}_j) \ ,
\label{eq:F}
\end{equation}
where
\begin{eqnarray}
{\bf k} & = & h {\bf b}_1 + k {\bf b}_2 + l {\bf b}_3 \\
{\bf x}_j & = & m_j^{(1)} {\bf a}_1 +  m_j^{(2)} {\bf a}_2
+ 
m_j^{(3)} {\bf a}_3 \ .
\label{eq:k_vector}
\end{eqnarray}
Here, the ${\bf a}_i$ are the crystal axes, the $m_j^{(i)}$ are the
crystallographic coordinates, and the ${\bf b}_i$ are the reciprocal
lattice vectors.  The $f_j({\bf k})$ are the form factors for the
given atomic species;\cite{IntCrystalC} the $o_j$ are the occupancy
numbers, which account for cell positions not always filled with an
atom or filled with atoms of different type with different
probabilities; and the $B_j$ are the isotropic Debye-Waller factors
that account for thermal vibrations in the lattice.

Our description of the contents of the cell is approximate, since the
oxygens and cations are excluded.  Also, since the structure of the
crystal is unknown, we do not have information about the occupancies
or Debye-Waller factors. Therefore, we set $o_j = 1$ and $B_j = 1/2$ in
all our trials.  These limitations imply that our calculated
diffraction pattern cannot exactly match the observed one even if we
locate perfectly all of the framework T-atoms.  Nonetheless, we can
capture the relevant features of the pattern. The contribution of the
oxygens to the diffraction pattern is less important than that of the
silicons, since $f_j$ is roughly proportional to the atomic
number. Indeed, it proves better to leave the oxygens out than to
include them at the midpoints between the T-atoms. The contribution of
non-framework species to the reflection intensities is suppressed
since they usually have large Debye-Waller factors, being more loosely
bound than the framework species, and often have fractional
occupancies.

The presence of multiple reflections at angles closer than the
resolution obtainable even with the best synchrotron radiation sources
is one of the major challenges to the use of powder data.  This
occurrence is common for zeolite samples.  In order to compare the
computed intensities with the experimental ones, we define a composite
peak at the average angle and place all of the intensity of the
multiple reflections into this composite peak.

To measure how well a particular configuration of T-atom positions can
match the experimental powder pattern, we define the quantity
\begin{equation}
H_{\rm PXD} = \frac{1}{N} \min_s \left[\frac{\sum_{i} { (I^{\rm obs}_i
- s I^{\rm calc}_i)^2/\omega_i}}{\sum_{i} 1/\omega_i}\right] \ ,
\label{eq:R}
\end{equation}
where $i$ runs over all of the $N$ peaks, which may be composites, the
$\omega_i$ are the weights, and $s$ is a global scaling factor.  We
make a similar definition for any available neutron data.  The
intensities are relative, and we scale the experimental intensities so
that the largest one is 1000.  The explicit expression for the global
scaling factor, $s_{\rm min}$, is
\begin{equation}
s_{\rm min} = \frac{\sum_i{ (I^{\rm obs}_i I^{\rm calc}_i)/ \omega_i}
 } {\sum_i{ \left(I^{\rm calc}_i \right)^2/\omega_i}} \ .
\label{eq:smin}
\end{equation}
The weights $\omega_i$ are associated to each peak according to the
following criterion: if the scaled intensity is less than 90,
$\omega_i = 1$; if $90 < I_i \leq 150$, $\omega_i = 2$; if $150 <
I_i \leq 300$, $\omega_i = 3$; and $\omega_i = 4$ otherwise.
These weights account for uncertainty both in the data and our fit to
the data, both roughly proportional to the intensity itself.

The numerical implementation of the calculation of the diffraction
terms can exploit the presence of the space group symmetry.  Since we
move one unique T-atom at a time, only the contribution of the unique
T-atom and of all of its symmetric images will change in $H_{\rm PXD}$
and $H_{\rm PND}$.  To calculate energy changes, only these terms need
to be reevaluated.  We can express the sum Eq.~(\ref{eq:F}) in the
following way:
\begin{equation}
F_{hkl} = \sum_\gamma F_{hkl}^{(\gamma)}\, ,
\;\;\;\;\;\;\;\;\; F_{hkl}^{(\gamma)} =
\sum_{j \in \gamma} f_j({\bf k}) \, o_j \, \exp(-B_j
{\bf k}^2 /4) \, \exp(i \,
{\bf k} \cdot {\bf x}_j) \ .
\label{eq:symF}
\end{equation}
Only one of the terms of the first sum changes when a single unique
T-atom is moved.

A good choice of the weights $\alpha_i$ in Eq.~(\ref{eq:figure}) is
crucial for the success of method.  The diffraction terms are very
sensitive to the positions of the atoms. In other words, they make for
a very rough energy profile.  Even a small change in the position of
one atom and its symmetric images can change $H_{\rm PXD}$ and $H_{\rm
PND}$ substantially.  The geometric potentials of Fig.~\ref{fig:pot}
are quite smooth. The density terms are also relatively smooth, even
when the move involves a change in the merging.  We want to mix the
geometric terms with the diffraction terms in such a way that the
roughness of the diffraction terms is smoothed out.  In order to find
the best values of these weights, we tried several different
combinations on one trial structure.  It was reassuring to find that
the rate of success was not sensitive to small changes in the
parameters.  A good choice for the $\alpha_{\rm PXD}$ or $\alpha_{\rm
PND}$ weight is between 1 and 2. The diffraction terms are extensive,
in the sense that the number of reflections is roughly proportional to
the total number of atoms, as are the other terms in the figure of
merit, but in few cases we found it necessary to increase or decrease
their importance with respect to the other terms.  The values of these
weights are $\alpha_{\rm T-T} = 1$, $\alpha_{\rm T-T-T} = 1$,
$\alpha_{\langle {\rm T-T-T} \rangle} = 2.0$, $\alpha_{\rm NB} = 1.5$,
and $\alpha_{\rm L} = 1$. The merging term $H_{\rm M}$ was in almost
all cases fixed to be zero at $r_{\rm M} = 0.8$~\AA\ and $-300$ at
$r_{\rm M} = 0$~\AA, and the associated weight was set at $\alpha_{\rm
M} = 1$.  The density weight $\alpha_{\rm D}$ was usually around $30$,
in order to ensure that the proper density was reached at low
temperatures.

\begin{figure}[p]
\epsfxsize=5in \centerline{ \epsfbox{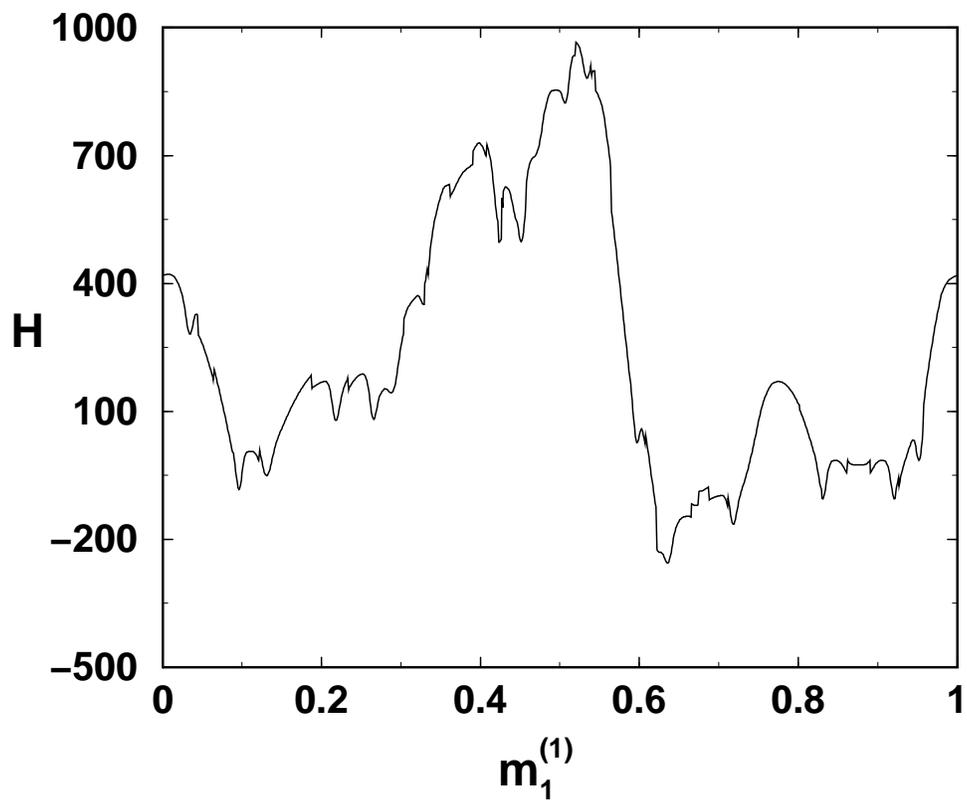}}
\caption{The figure of merit profile as a function of one crystallographic
coordinate, $m_1^{(1)}$, for the faujasite framework.}
\label{fig:profile}
\end{figure}

How the figure of merit depends on the atom coordinates is shown in
Fig.~\ref{fig:profile}.  The curve corresponds to a one dimensional
slice of the full profile, obtained by sliding one coordinate of one
T-atom across the unit cell of the zeolite faujasite.  The profile is
very rough, with many narrow valleys. It is crucial to notice that the
position of the atom in the faujasite structure corresponds to the
global minimum of the curve, at roughly $m^{(1)}_1 \simeq 0.6$. If we
were solving this material, this position would be the minimum that we
would have to locate in order to solve the structure. Clearly the
figure of merit is very rough, and we will need a powerful simulation
protocol to perform the many-dimensional global optimization.

The figure of merit defined by Eq.~(\ref{eq:figure}) may possess
invariance under translations of the unit cell in particular
directions.  This reflects the fact that for some space group
symmetries the position of the unit call may be arbitrary.  This is
clear, for instance, in the case of the space group P1.  In this
space group, the unit cell can be moved freely in all directions, without
changing the description of the system.  Our figure of merit would be
invariant under a simultaneous, arbitrary, and continuous translation
of the atoms.  The space groups that have this type of freedom are
called polar, and the directions in which the unit cell can be moved
are the polar directions.  In other instances, the atoms within the
unit cell can be moved a discrete amount in one direction, such
one-half the unit cell, and still lead to the same crystal structure.
In both of these cases it is useful to eliminate these translationally
invariant modes.  It is straightforward to identify a polar group and
the polar directions and to define a projection operator that will
restrict proposed moves to an orthogonal subspace of the polar
directions.  It is enough to select one atom that can move only along
the orthogonal subspace to break the polar symmetry.  In the case of
unit cells with two or more choices of cell setting, such as occurs in
the framework LTL, the ambiguity can be eliminated by choosing an
appropriate asymmetric unit and limiting the movement of the unique
T-atom to that cell.  Use of an asymmetric unit, however, was found
not to be necessary for the success of the method.
\section{The Monte Carlo}
\label{sec:montecarlo}
In this section we will describe in some detail the Monte Carlo
algorithm that we use to sample the figure of merit.  We will first
make some general comments about biased Monte Carlo importance
sampling. We then describe simulated annealing and parallel tempering.

Monte Carlo methods have been used extensively since their
inception\cite{Metropolis} to sample equilibrium probability
distributions of systems with many degrees of freedom.  The key step
is to define a Markov process that evolves the system from
configuration to configuration.  As long as this Markov process
satisfies certain properties, one is assured that after $N$ steps,
time averages will approximate ensemble averages to within a relative
error of $1/\sqrt{N}$.  Specifically, if the Markov process is ergodic
and regular and satisfies detailed balance, it can be shown that the
limiting probability distribution is the one we seek.  The proof uses
the Perron-Frobenius theorem and the fact that a matrix obeying
detailed balance has a complete set of eigenvectors (see, for example,
Ref.~\onlinecite{ParisiBook}).  A more general proof shows that the
method need satisfy only the weaker balance condition.

One of the shortcomings of the traditional Metropolis method is that
it does not use any information about the energy landscape around the
current configuration when picking trial moves.  Oftentimes, the
proposed move brings the system to regions of configuration space that
are high in energy, and the move is rejected.  These rejected moves
hinder effective sampling of the Boltzmann distribution.

\begin{figure}[p]
\epsfxsize=5in \centerline{ \epsfbox{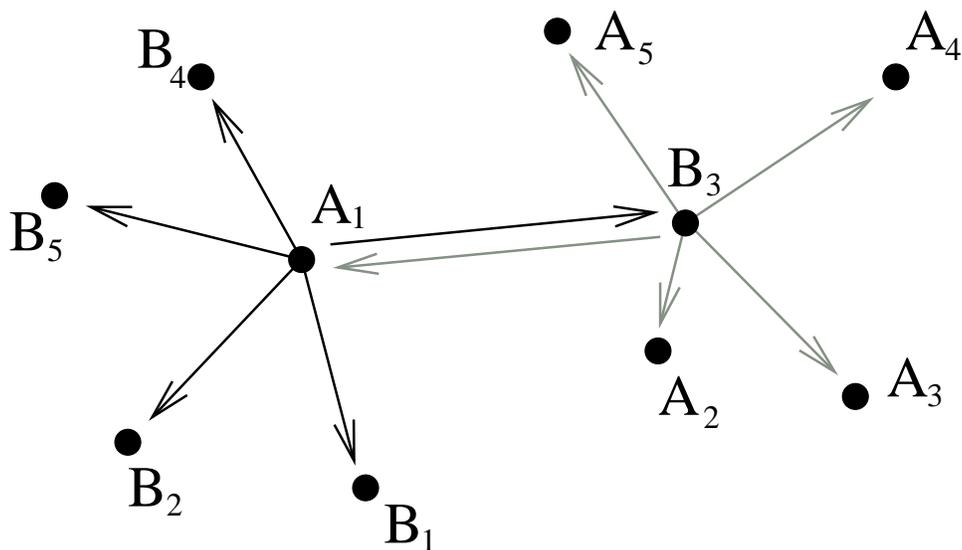}}
\caption{The biased displacement move. In this case $k=5$, and $B_n = B_3$. The
arrows represent the transition probabilities.}
\label{fig:bias}
\end{figure}

Biased Monte Carlo methods have been shown to improve sampling in many
cases.  They were originally introduced to lead to more efficient
simulations of complex liquids.\cite{II_Frenkel,II_SmitV,II_dePablo}
The basic idea is to probe the configurations around the current one
and to {\em propose} moves that are more likely to be accepted.  In
our case the biased move proceeds as follows.  Let us call the current
configuration $A_1$.  We extract $k$ random displacements, $\Delta
{\bf x}_i$, which define $k$ proposed new configurations,
$B_i$. These moves are extracted from a Gaussian distribution
\begin{equation}
p^{\rm int}_i = \frac{\exp[-\Delta {\bf x}_i^2/(2\sigma^2)]}
  		{[2 \pi \sigma^2]^{3/2}} \ .
\label{eq:p_int}
\end{equation}
We construct the Rosenbluth weight $W$, defined as
\begin{equation}
    W(n) = \sum_{i=1}^k \exp[-\beta H(B_i)] \ ,
\label{eq:rosen}
\end{equation}
and we assign a normalized probability 
\begin{equation}
    p^{\rm ext}_i = \exp[-\beta H(B_i)]/W(n)
\label{eq:probability}
\end{equation}
to each configuration $B_i$ (see Fig.~\ref{fig:bias}).  We randomly
select one of these configurations, $B_n$, according to its
probability.  The configuration $B_n$ is our proposed move.  Clearly
the lower the energy is, the more likely the configuration will be
selected. In order to satisfy detailed balance, we must modify the
acceptance probability of the proposed move.  This requires us to
calculate the likelihood of the reverse move $B_n~\to~A_1$.  The super
detailed balance condition, which ensures detailed balance, can be
satisfied by defining a set of $k - 1$ new trial moves, $A_j$, from the
proposed configuration, $B_n$.\cite{FrenkelSmit}  The set $\{A_1,
A_j\}$ defines the reverse Rosenbluth weight
\begin{equation}
    W(o) = \exp[-\beta H(A_1)] + \sum_{j=2}^{k} \exp[-\beta H(A_j)] \ ,
\label{eq:rosen2}
\end{equation}
and the normalized probability of selecting the reverse move is
\begin{equation}
    p^{\rm ext}_o = \exp[-\beta H(A_1)]/W(o) \ .
\label{eq:oldprobability}
\end{equation}
The super detailed balance condition can now be written as
\begin{equation}
\pi(A_1) \; T(A_1 \to B_n)\; {\rm acc}(A_1 \to B_n) = 
\pi(B_n) \; T(B_n \to A_1)\; {\rm acc}(B_n \to A_1) \ ,
\label{eq:detailbiased}
\end{equation}
where $\pi(A) \propto \exp[-\beta H(A)]$ is the limiting distribution
that we want to sample, and the probability of accepting the proposed
move is acc$(A_1 \to B_n)$. The forward transition probability $T(A
\to B_n)$ is just the probability of selecting the configuration,
$p^{\rm int}_n p^{\rm ext}_n$, and the reverse transition probability
is $p^{\rm int}_o p^{\rm ext}_o$.  The super detailed balance
condition is, then,
\begin{equation}
    \frac{{\rm acc}(A_1 \to B_n)}{{\rm acc}(B_n \to A_1)} =
	\frac{\pi(B_n)}{\pi(A_1)}
	\,\frac{p^{\rm int}_o}{p^{\rm int}_n}
	\frac{p^{\rm ext}_o}{p^{\rm ext}_n} = \frac{W(n)}{W(o)} \ .
\label{eq:acceptbias}
\end{equation}
A reasonable choice for the acceptance probability is
\begin{equation}
   {\rm acc}(A_1 \to B_n) = \min\left(1, \frac{W(n)}{W(o)}\right) \ .
\label{eq:minaccept}
\end{equation}
This class of biased moves significantly improves the sampling of our
algorithm with respect to the simple Metropolis scheme.  We found that
a biased move with $k = 5$ works well.

The figure of merit described in the previous section gives a
quantitative measure of how well a particular arrangement of atoms
resembles a zeolite.  We are interested in minimizing the figure of
merit in order to find the most reasonable arrangements.  Biased Monte
Carlo alone is unable to sample efficiently the rough figure of merit
at low temperatures. Sampling can often be achieved, however, with
simulated annealing.\cite{kirkpatrick83,kirkpatrick84} In this
approach, a series of simulations at progressively lower temperatures
is performed, and the distribution at each temperature is sampled
using a Monte Carlo method, with or without biasing.  The simulation
is started from a high temperature and the temperature is
progressively reduced according to an annealing scheme.  In general
the temperature is kept unchanged for a fixed number, $N$, of Monte
Carlo steps.  After these $N$ steps the temperature is reduced
according to $T^\prime = \kappa T$, where $\kappa < 1$, and this cycle
is repeated until the temperature is such that most Monte Carlo moves
are rejected, and the system is effectively frozen.

The width of the distribution of proposed moves, $\sigma$ in
Eq.~(\ref{eq:p_int}), can be adjusted during the simulated annealing
run.  In general, for a fixed value of the temperature one chooses
$\sigma$ so that a reasonable number of moves are accepted. We call
the ratio of accepted to attempted moves the acceptance ratio $g$. On
the one hand, if the trial moves are small, then most moves are
accepted, $g \simeq 1$, but the effect on the energy is minimal, and
the probability distribution is sampled ineffectively.  On the other
hand, if the trial moves are large, then most moves are rejected,
and the sampling is also ineffective.  In simulated annealing the
issue of acceptance is further complicated by the fact that the
temperature is changed.  We found it convenient to fix a target
acceptance ratio, $g_{\rm t}$, and to adjust the size of the proposed
move distribution, $\sigma$, so as to make the actual $g \simeq g_{\rm
t}$.  In general we start with a large width, $\sigma = 3$~\AA, and we
lower it during the annealing to values around $\sigma \simeq
0.5-1.0$~\AA\ at low temperature.  We use a proportional control
scheme
\begin{equation}
 \sigma^\prime = \sigma [1 + \epsilon (g - g_{\rm t})] \ ,
\label{eq:proportional}
\end{equation}
to adjust $\sigma$ each time we reduce the temperature.  Even though
$\sigma$ lags with respect to the temperature, since we use the $g$
measured at the higher temperature, we found this control scheme
to be effective.

\begin{figure}[p]
\epsfxsize=5in \centerline{ \epsfbox{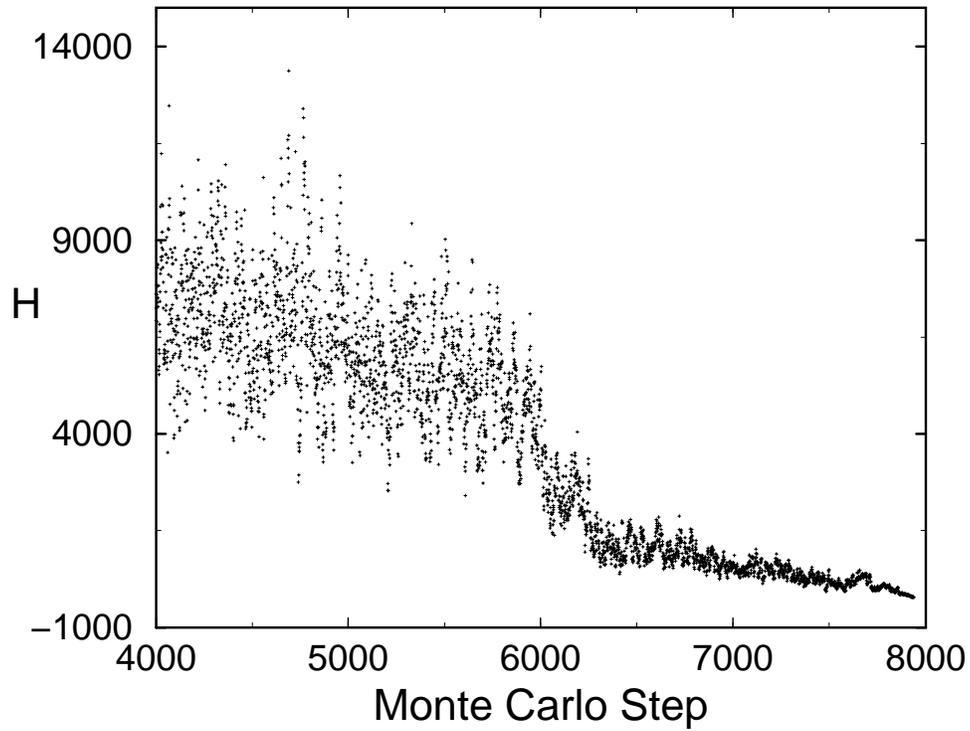}}
\caption{An energy trace for a simulated annealing run.  Only the
portion relative to the annealing phase is shown.}
\label{fig:SA_ene}
\end{figure}

The initial high temperature is found by fixing $\sigma$ and
performing short trial runs with a Metropolis Monte Carlo. We always
start from a completely random initial condition. The temperature is
doubled until the fraction of accepted moves during the trial run
exceeds a given threshold value.  We found that a threshold of $g =
0.5$ is always sufficient to get to a high enough temperature.  Once
this initial stage is completed, we thermalize the structure at this
fixed temperature with the biased moves.  This ensures that we lose
track of the initial condition.  We then start cooling the system
according to a preset annealing schedule.  In all but a few cases we
used $N = 200$ and $\kappa = 0.8$.  A typical annealing energy trace
is shown in Fig.~\ref{fig:SA_ene}.

We will show in Section~\ref{sec:results} how effective the
combination of simulated annealing and biased moves is in finding the
correct frameworks of known zeolites.  Typically one or a few runs at
most are needed to solve a structure.  If the first run is not
successful, we try again with different initial positions and random
seed, and eventually the correct structure is found.  For complex
structures with many unique atoms, $n_{\rm unique} \geq 8$, this
approach sometimes fails to converge to the correct structure within a
reasonable time. Of course, one could try to use different annealing
schedules, or try with a greater number of different initial
conditions.  In principle, nothing prevents this method from finding
the correct solution.  Nonetheless, slow equilibration is more than
just a technical detail.  For one, during simulated annealing the
system is not at equilibrium, since the temperature is reduced at
regular intervals.  For another, once the system falls in a local
minimum in the rough energy profile (Fig.~\ref{fig:profile}), and the
temperature is too low for the system to escape via fluctuations in a
finite number of steps, the system is stuck.

One other method stands out as a good candidate for sampling
probability distributions with complicated landscapes: parallel
tempering.  This method was developed as an effective Monte Carlo
procedure for the study of systems with large free
energy barriers.\cite{geyer91,geyer95} This method was later applied to
spin glasses,\cite{hukushima96,marinari98} self-avoiding random
walks,\cite{tesi96,tesi96b} lattice QCD,\cite{Boyd98} and studies of
biological molecules.\cite{hansmann97} Following
Ref.~\onlinecite{marinari98}, we call the method parallel tempering,
for its similarity to simulated tempering, a related method also
proposed as an improvement on simulated annealing.\cite{marinari92}
J-walking is a similar method,\cite{frantz90} often used in molecular
optimization problems.  J-walking is not an exact Monte Carlo
scheme,\cite{geyer95,frenkel96:page181} due to the non-Markovian reuse
of configurations.  While none of the implementations of parallel
tempering cited above is Markovian at the level of a single move, it
is a simple matter to make such an implementation.  Indeed, taking
care with the definitions allows one to understand how to optimize the
parallel tempering method by more frequently updating the systems at
lower temperature and with longer autocorrelation times.

\begin{figure}[p]
\epsfxsize=5in \centerline{ \epsfbox{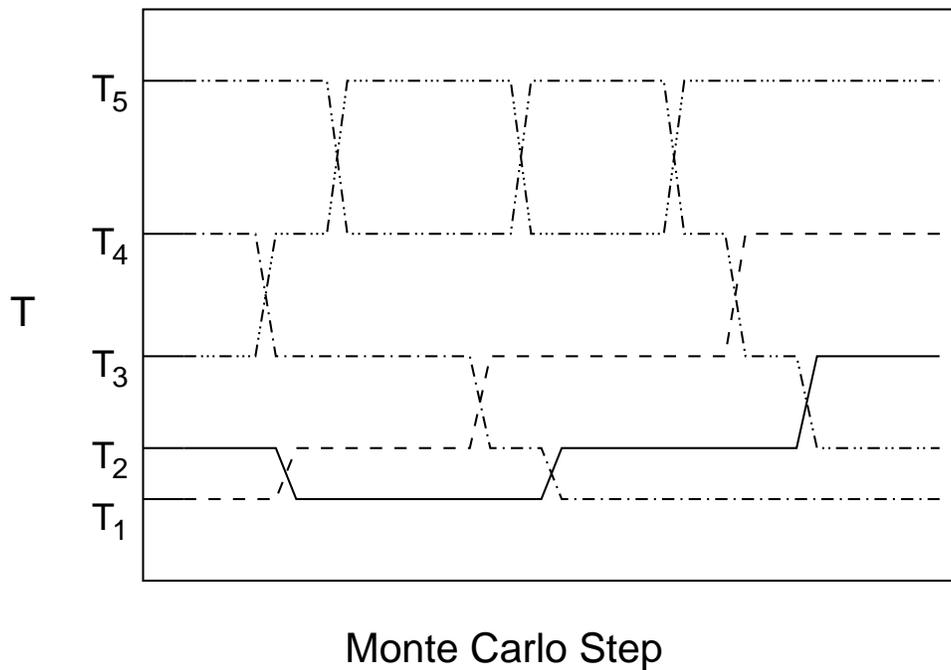}}
\caption{A schematic drawing of the swapping taking place during a
parallel tempering simulation.}
\label{fig:tempering}
\end{figure}

The idea of parallel tempering is to consider $n$ systems, each in a
canonical ensemble, and each at a different temperature.  We define
the instantaneous configuration of system $i$ at Monte Carlo step $t$
to be $C_i(t)$. Each system $i$ has a different temperature $T_1 < T_2
< \ldots < T_n$, where $T_1$ is the low temperature that we want to
sample, and $T_2, \ldots, T_n$ are higher temperature systems that aid
in the sampling.  The extended canonical ensemble is given by
\begin{equation}
{\cal Q}  = \prod_{i=1}^n {\cal Q}_i \ ,
\label{eq:part_pt}
\end{equation}
where ${\cal Q}_i$ is the canonical partition function,
\begin{equation}
{\cal Q}_i = \sum_{\{C_i\}} \exp[-\beta_i H(C_i)] \ .
\label{eq:part}
\end{equation}
We introduce a swap move, which proposes the exchange of two copies at
different temperatures.  The proposed move is accepted according to
the Metropolis rule.  We compute the action difference that the swap
move introduces,
\begin{equation}
\Delta S =  \beta_j H(C_i) + \beta_i H(C_j) - \beta_j H(C_j) -
\beta_i H(C_i) = (\beta_j - \beta_i) [H(C_i) - H(C_j)] \ .
\label{eq:swapdelta}
\end{equation}
To satisfy detailed balance, we accept the move with probability
\begin{equation}
p = \min[1, \exp(-\Delta S)] \ .
\label{eq:accept}
\end{equation}
Typically we consider swaps between adjacent temperatures, $j = i +
1$.  For a good choice of temperatures, swaps will be accepted with a
significant probability.  We show in Fig.~\ref{fig:tempering} a
schematic drawing of the swapping process.

Parallel tempering allows the system to escape local minima by
swapping with the systems at higher temperature.  The choice of
temperatures should be such that the high temperature, $T_n$, is great
enough so that the extended ensemble can effectively surmount the free
energy barriers.  The intermediate temperatures create a ladder that
the system uses to climb over the barriers.  It is important to notice
that the extended ensemble is precisely defined by
Eq.~(\ref{eq:part_pt}).  This means, for example, that system $i$
samples the canonical ensemble at temperature $T_i$. We satisfy
detailed balance because of Eq.~(\ref{eq:accept}).  The displacement
and swapping moves are clearly ergodic, in principle. If our moves are
defined so as to produce a Markov process, then we are guaranteed to
sample the extended ensemble in Eq.~(\ref{eq:part_pt}).

\begin{figure}[p]
\epsfxsize=5in \centerline{ \epsfbox{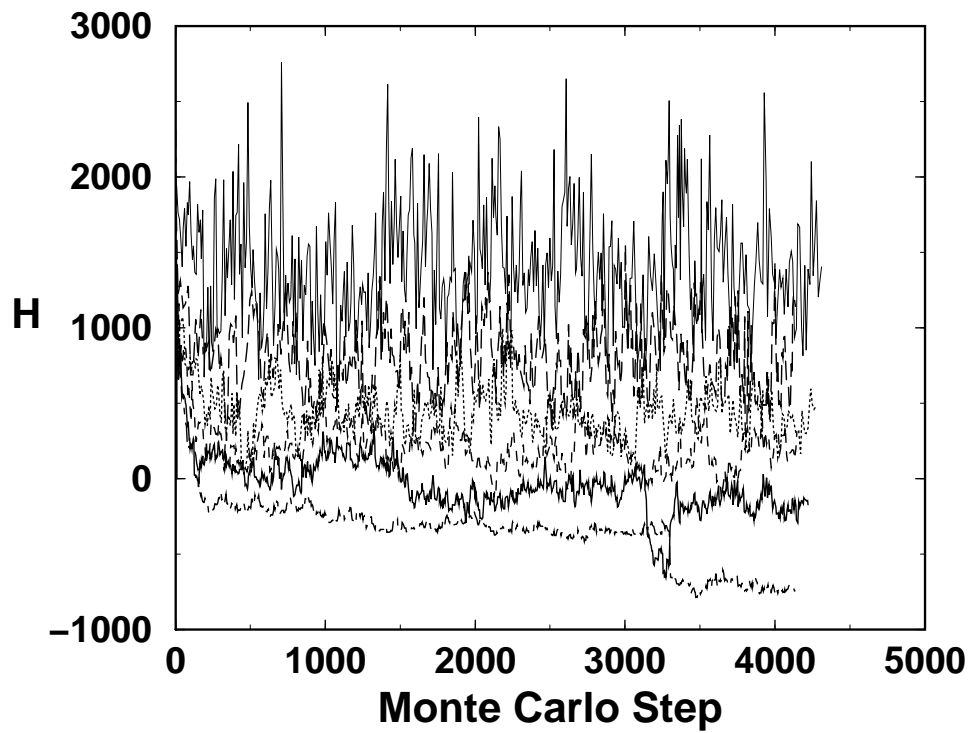}}
\caption{Traces of the energy in a parallel tempering run on MFI.
 \hspace{3in}}
\label{fig:pt_traces}
\end{figure}

When there is more than one kind of update rule in a Monte Carlo
simulation, the moves must be selected {\em randomly} in order to have
a Markov chain on the level of a single move.\cite{frenkel96:page50}
Of course, one is free to pick the relative probabilities of selecting
each type of move.  Our implementation of parallel tempering selects
the moves at random.  We start by selecting one of the systems at
random.  We then randomly decide whether to make a swap move or a
displacement move. We have found that choosing a displacement move
90\% of the time leads to efficient sampling.  If a displacement move
is selected, one of the unique atoms, chosen at random, is updated.
If a swap move is selected, we attempt to swap the chosen system with
the system at higher temperature.  Since the systems at low
temperature are slower to evolve under the Monte Carlo sampling, we
pick these systems more frequently than the ones at higher
temperature.  We typically allow the two lowest temperatures to be
updated twice as frequently, leading to more swapping and more
displacement moves for these systems. In the general case, the update
frequencies should be proportional to the autocorrelation times of the
respective systems, as measured in the parallel tempering
simulation. Each temperature also has an associated, constant move
amplitude $\sigma_i$ that is adjusted at the onset in order to have a
reasonable acceptance ratio.  We note that the swap move is very fast
in computing time, since the current energy of each configuration is
stored.  The initial conditions are $n$ random arrangements of atoms.
We do not thermalize the systems, since the parallel tempering swaps
are very effective in arranging the configurations according to their
energies. A typical energy trace for a parallel tempering run is shown
in Fig.~\ref{fig:pt_traces}.

In general, one may be interested in averages of all the systems at
the various temperatures, $T_i$.  In this case, parallel tempering can
be used to study phase diagrams, and the additional equilibration
given by the swapping helps the systems at low temperature to sample
the probability distribution effectively.  We are specifically
interested in the system with the lowest temperature, which will hop
between likely zeolite structures until the one corresponding to the
experimental sample is found.  We monitor the structure at lowest
energy, and we stop when all the atoms are 4-coordinated and the
diffraction term indicates that a good match has been found.

\begin{figure}[p]
\epsfxsize=5in \centerline{ \epsfbox{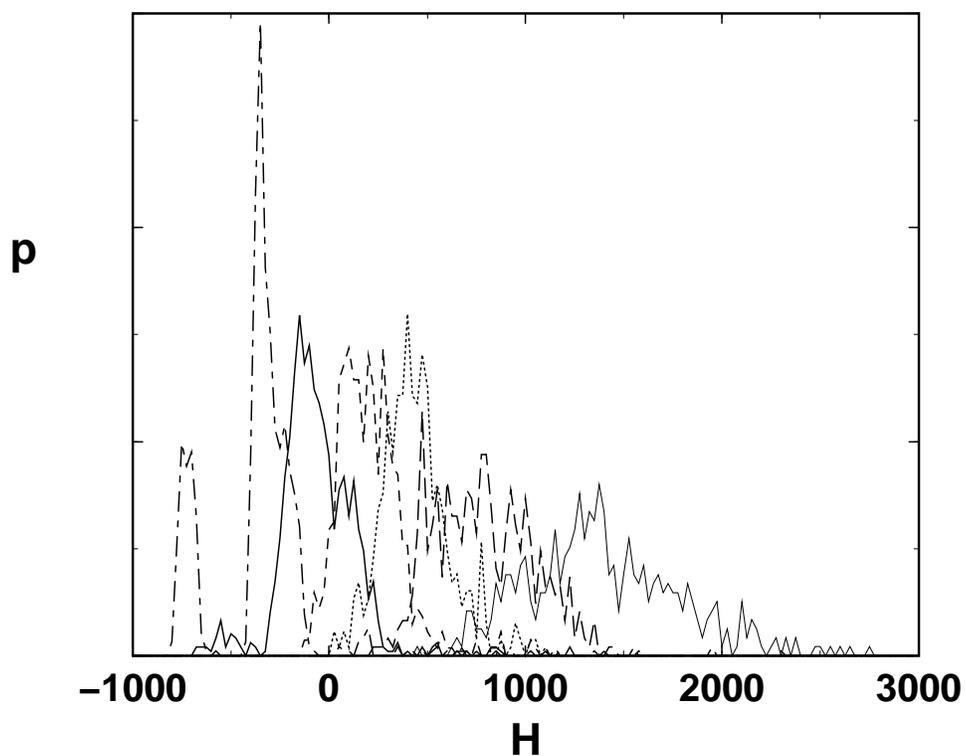}}
\caption{Histograms of the energies observed in a parallel tempering
run on MFI. Note the overlap of the distributions.}
\label{fig:histo}
\end{figure}

The choice of temperatures is very important in parallel tempering.
To determine them, it is useful to consider the energy fluctuations.
We plot the energy histograms of the Monte Carlo data for a parallel
tempering run, and we construct the temperature ladder so that the
histograms overlap significantly.  A typical example is shown in
Fig.~\ref{fig:histo}.  A good choice for the temperatures can usually
be obtained from an initial simulated annealing run.  This allows us
to locate the freezing temperature, a high temperature, and a low
temperature.  If necessary, the temperature selection can be refined
to ensure that the energy histograms overlap and that the copies
traverse the temperature ladder in the parallel tempering.  This
criterion leads to a ladder of temperatures that is automatically
inferred from the properties of the system, rather than guessed or
estimated. From Fig.~\ref{fig:histo}, for example, it is clear that
the high temperatures can be spaced more widely than can the low
temperatures.
\section{Results}
\label{sec:results}
To assess the usefulness of this structure solution method we test it
on the 118 publicly known zeolite structure types.  For each of these
materials, the chemical composition and the atomic positions are
known.\cite{atlas_book,III_NewsamI,IZA_web} We describe here how the
method fares on these known materials.

Since we do not have actual experimental diffraction information for
most of these materials, we use the available data to construct
synthetic X-ray diffraction patterns.  The available data include
information about the unit cell size and parameters, the space group
symmetry, the type of atoms present in the unit cell, the atomic
positions, the occupancies, and the Debye-Waller factors.  The
reflections included in our diffraction pattern are the ones in the
range $5^{\rm o} \leq 2\theta < 35^{\rm o}$. This excludes the low
angle data, which usually have background contributions, and the high
angle data, which are usually poorly resolved. The multiplicities of
the peaks are accounted for in the production of the powder
patterns. We use a wavelength of $\lambda = 1.54056$ \AA.  We assume a
crystallite size of $1 \mu$m, which is typical for new zeolite
samples. We assume that the data could be collected on synchrotron,
and so we use instrument parameters appropriate for a beam line.  With
these assumptions, the diffraction pattern has a peak resolution of
approximately $\Delta (2 \theta) = 0.06$.  We use this relatively
conservative criterion when forming the composite peaks. Of course,
the multiplicities of the peaks were accounted for in the production
of the powder patterns.  We verified that we could reproduce the
intensities obtained with Cerius2, such as the ones in
Fig.~\ref{fig:pxd}. In the cases where the framework topology can be
described by a higher symmetry, we often, but not always, use the
higher symmetry setting for the solution.  This reduces the number of
degrees of freedom, making the solution simpler and faster.  As
before, $n_{\rm unique}$ is the number of crystallographically
distinct T-atoms used in the simulated annealing, and $n_{\rm max}$ is
the number of crystallographically distinct T-atoms in the maximal
symmetry setting.  In the cases where several zeolite samples are
available for a given framework, we generate the diffraction pattern
using the material with the highest silicon content, or the lowest
number of non-framework species.

The result of applying the solution procedure to the known zeolites is
shown in Tables~\ref{tbl:res1} and \ref{tbl:res2}. For each framework,
we attempted a simulated annealing run with biased Monte Carlo moves.
At the end of the run, we computed the coordination sequence of the
unique atoms.  The coordination sequence is a list of integers that
counts the number of neighbors one, two, and so on connections
away.\cite{meyer79}  It uniquely identifies a given structure through
its topology, rather than through the precise locations of the atoms.

In the cases where we were unable to solve the structure in a few
simulated annealing runs, we turned to parallel tempering.  Using the
energy histograms collected at the various temperatures in the
simulated annealing run, we set up a ladder of five or six
temperatures, using the corresponding move amplitudes, $\sigma_i$, from the
simulated annealing.  Using parallel tempering we were able to solve
all the structures that we attempted.  Depending upon the complexity
of the structure, which is roughly proportional to the number of atoms
in the unit cell, a solution is achieved in 0.2--4 hours on a Silicon
Graphics Indigo$^2$ with a 195 MHz R10000 processor.

\section{Discussion}
\label{sec:discuss}
From the results shown in Tables~\ref{tbl:res1} and \ref{tbl:res2} we
see that the introduction of biased moves in the simulated annealing
dramatically improve the success rate of the method.  The $N_{\rm MC}$
column refers to the number of simulated annealing attempts needed to
solve the structure with simple Metropolis Monte Carlo moves, while
$N_{\rm BMC}$ refers to the number of attempts required with biased
moves.  It is apparent that in most cases the biased moves
substantially improve the sampling, allowing one to find the correct
structure in fewer trials.  More importantly, the technical
limitations encountered with Metropolis Monte Carlo\cite{Deem92a} have
been mostly removed, since most structures not solvable with simple
moves can be solved with biased moves.  It must be noted that several
of the structures shown in the Tables were not known at the time
Ref.~\onlinecite{Deem92a} was published, and, arguably, some of the
simpler new structures could have been solved with Metropolis moves.
Indeed, at least four groups have used the Metropolis Monte Carlo
approach,\cite{Deem92a} implemented in Cerius2, to solve new zeolite
structures.\cite{Akporiaye96a,Akporiayeb,Akporiaye96c,Stucky93,Cheetham,Vaughan}
Zeolite beta provides an example of a well known and important zeolite
that can be solved with the biased Monte Carlo, but not with
Metropolis Monte Carlo.  The structures that were not solved with
simulated annealing, and only these, were attempted with parallel
tempering. All were solved.  From the success of this approach, it is
clear that parallel tempering is a much more powerful method than
simulated annealing.  We believe that parallel tempering is a much
better method precisely because it samples the correct equilibrium
distribution.

The most complex of the publicly known zeolites is ZSM-5. The
framework structure type is MFI, and it is shown in
Fig.~\ref{fig:mfi}.  This structure is the most complex because it has
the highest number of unique T-atoms, $n_{\rm unique} = 12$.  We were
able to solve this structure in one day's work that included the
simulated annealing run, selecting the parallel tempering
temperatures, and performing the actual parallel tempering run.  The
energy traces for the latter are shown in Fig.~\ref{fig:pt_traces},
and one can clearly see how the correct structure is found in one of
the high temperature systems and then swapped down to the lowest
temperature system.  It is also clear from the trace that without the
swapping, the lowest temperature system would never cross the energy
barriers separating the initial condition from the correct structure.
To showcase the power of the method, we solved solved NU-87, framework
code NES, in a low symmetry setting with 17 unique atoms.  This proves
that parallel tempering is powerful enough that the use of the maximal
symmetry setting is not required.  Although an upper limit to the
practical applicability of this method must exist, that limit is not
obvious from the results of our trial runs.

We encountered a few types of problems when testing our method on 
the known zeolites.
  In some cases the framework
has loops of length 3, with 3 T-atoms connected in a triangle.
Clearly the bond angles in this case are far from the usual tetrahedral
value, and the potentials that we defined in Section~\ref{sec:fom}
may be incorrect for these particular structures.  In other words, the
correct structure may not correspond to minimum of the geometric
terms in the figure of merit.  In these cases, we found that a small
adjustment of the geometric weights, $\alpha_i$ in
Eq.~(\ref{eq:figure}), is a sufficient remedy.  These cases can be
recognized by a visual inspection of the structures produced,
 even with no {\em a
priori} knowledge of the correct structure.   The case of
open frameworks, frameworks for which not all of the T-atoms are
4-coordinated, can be treated by not penalizing 3-coordinated atoms.
These frameworks are listed in Table~\ref{tbl:not}.

A few structures were harder to solve because of their merging.  This
is the case, for example, for MEP, PAU, and MWW.  Let us consider MWW,
which has $n_{\rm unique} = 8$, $n_{\rm symm} = 24$, and $n_{\rm o} =
72$.  The difficulty in this case stems from the fact that two
combinations of merged atoms give the same $n_{\rm T}$: ($12 \times 5$
atoms + $4 \times 3$ atoms) $= 72$ atoms and ($12 \times 4$ atoms + $6
\times 4$ atoms) $= 72$ atoms. Many more combinations give numbers
close to 72.  The density term, $H_{\rm D}$, of Eq.~(\ref{eq:figure})
will not distinguish between these combinations.  This occurrence can
be inferred by visually inspecting the configurations of minimum
energy generated or by tabulating the number of merged atoms
associated with each unique atom.  Again, no knowledge of the correct
structure is needed to realize that the figure of merit has two or
more deep wells separated by high barriers.  The case of MWW is
further complicated by the presence of 3-loops and the presence of two
atoms which, although not connected, are separated by just 3.6 \AA.
In the case of MWW only, we adjusted the parameters $\alpha_i$
slightly, making $\alpha_{\rm D}$ small and reducing the range of the
interactions. This allowed us to solve the structure with 3 iterations
of parallel tempering. Because of these geometric irregularities, MWW
was actually the most difficult structure for us to solve, even though
it has only 8 unique T-atoms.

The presence of template molecules or heavy cations in the structure
may cause a greater problem.  In this case, the $H_{\rm PXD}$ term
favors the presence of a nonzero scattering density in the regions occupied by
the non-framework species.  This makes the diffraction term ambiguous,
and one can find many incorrect structures that are still feasible
from the point of view of geometry alone.  The preferred solution in
this case is to calcine the structure to remove the template or to
exchange the heavy cations for lighter ones.  If the material is
not stable to this treatment, then the non-framework species can be
included as degrees of freedom in the figure of merit.  

Zeolites with metal substitutions are quite common.  Several of the
structures that we solved are aluminophosphates or contain gallium,
beryllium, cobalt, or zinc along with phosphorus.  The effect of
substitutions is quite dramatic, both on the geometry and on the
diffraction.  In fact, preferred bond lengths may change.  The fact
that the atomic species are different usually lowers the maximal
symmetry allowed, and hence increases the number of unique atoms.  We
solved all known instances of these materials using just silicon to
match the diffraction data.  When possible, we used the higher
symmetry allowed by assuming that all T-atoms are identical in the
structure determination.  Of course, it would be straightforward to
simulate frameworks with different T-atoms and to include an atom
exchange move in the Monte Carlo.  This move, selected at random,
would be accepted with a Metropolis criterion, and would lead to
structures with the correct framework and the correct atomic species
in each position.  The case of the framework WEI, weinebeneite, is
separate since this is a beryllium phosphate, and the bond lengths are
quite different from regular aluminosilicates. A simple redefinition
of the geometric terms in Fig.~\ref{fig:pot} would allow us easily to
solve this material.  A general extension would be to allow for
species with with different coordination numbers as well as different
T--T distance and T--T--T angle potentials.  This extension would
broaden the range of applicability of the method well beyond zeolites.

The issues of thermalization, autocorrelation times, and efficiency
are usually addressed in numerical simulation studies.  In our case,
we are concerned with these elements in a qualitative sense only.  In
fact, all that matters for the utility of our method is that it
determine the structure of an unknown material in a reasonable amount
of time.  We have shown that this time is very reasonable and very
small compared to the time it takes to synthesize a new zeolite. The
time taken by this approach should be compared to the time it takes to
solve a single structure with conventional methods, typically measured
in weeks, months, or year, or years.  Timing issues may become
relevant when one is interested in exhaustively listing all of
the possible low energy geometries for a given cell size and symmetry,
without the aid of experimental data.  Hypothetical structures of this
sort could help, for example, in the {\em design} of new zeolite
materials, when the synthesis process is better understood.

\section{Conclusion}
\label{sec:conclusion}  
We have introduced a powerful biased Monte Carlo approach that can
determine the structure of a new zeolitic material from the powder
diffraction pattern and density, both of which are easily measured in
experiments.  All of the publicly known zeolites were solved in a
realistic test application of the method.  The method is rapid and
automatic, making it a natural tool for use within the combinational
chemistry paradigm.  The proposed technique can also be used to
generate hypothetical zeolite structures.  The number of structures
that can be generated in this way far exceeds the number constructed
to date by hand.\cite{III_Smith} A database of structures can be built
that will allow synthetic chemists to search for a structure that
matches the X-ray powder diffraction pattern of newly made materials.
The generation capability may also be directed, by, for example,
specifying large pores, low-densities, or other interesting functional
properties.  Such use allows the creation of hypothetical zeolites
with new, tailored structures.  These structures can then be sought in
rational syntheses.\cite{III_Davis} The direct, real space approach
may also be applied to small, low quality molecular
crystals,\cite{III_DeemVII} such as those from drugs, dyes, pigments,
and organic non-linear optical materials, for which only powder
diffraction data are available.

Perhaps of more general interest to the simulation community is the
value of our work as a case study on the effectiveness of parallel
tempering.  We have shown that parallel tempering, combined with
biased Monte Carlo, is a powerful method for molecular systems with
many local minima.  We have suggested a general and simple histogram
method to determine the temperatures required in the extended
ensemble.  By considering how the energy histograms overlap, we
generally expect that, away from critical points, the number of
required temperatures is proportional to $[E_{\rm mix}(N) - E_0(N)] /
\sqrt N$, where $N$ is the number of degrees of freedom, $E_0$ is the
energy at the temperature of interest, and $E_{\rm mix}$ is the energy
at which the system can overcome all relevant barriers.  How the
number of required temperatures scales with system size will depend
entirely on how $E_{\rm mix}$ scales with $N$.  In simple cases,
$E_{\rm mix}$ may scale as a correlation volume.  In the worst case,
$E_{\rm mix}$ will scale as $N$, and so the number of temperatures
will be proportional to $\sqrt N$.  By formulating the method as a
Markov process on the level of a single move, we were able to provide
a general and more efficient strategy for the choice of the individual
system update frequencies.  Based upon our experience with zeolite
structure solution, we recommend use of parallel tempering whenever
simulated annealing struggles on a minimization problem, especially if
correct sampling of a low-temperature distribution is desired.

\acknowledgments 
It is a pleasure to acknowledge useful discussions with John Newsam
and Giorgio Parisi.  MF acknowledges useful discussions with Gerard
Jungman.  This research was supported by the National Science
Foundation through grant CTS--9702403, by the University of California
Energy Institute, by the Donors of the Petroleum Research Fund, and by
Molecular Simulations, Inc.

\bibliography{zeolite}
\pagestyle{empty}
\renewcommand{\arraystretch}{0.51}
\begin{table}[htbp]
\caption{The weights of T-atoms according to the number of
	 neighbors.}
\label{tbl:bond}
\begin{center}
\begin{tabular}[t]{cc} 
Neighbors & Weight \\ \hline
       0 & 1000\\
       1 & 650\\
       2 & 300\\
       3 & 100  \\
$\geq 4$ & 0 \\ 
\end{tabular}
\end{center}
\end{table}

\begin{table}[htbp]
\caption{Results for zeolite frameworks A to G.\protect\tablenote{
The space group and the number $n_{\rm unique}$ of
crystallographically distinct T-atoms used in the structure solution
is listed for each zeolite framework.  The number of distinct T-atoms
in the maximal symmetry setting is $n_{\rm max}$.  The number of
symmetry operators in the chosen setting is $n_{\rm symm}$.  The total
number of T-atoms in the unit cell is $n_{\rm T}$.  The number of runs
required to solve a given structure with Metropolis Monte Carlo and
simulated annealing is $N_{\rm MC}$, with a dash indicating no
solution found.  Each run used the same input parameters and
differered only in the initial random number seed.  Similarly, the
number of runs required to solve a given structure with biased Monte
Carlo and simulated annealing is $N_{\rm BMC}$.  Finally, $N_{\rm PT}$
is the number of runs required to solve via biased Monte Carlo and
parallel tempering those structures not solved with simulated
annealing.}}
\label{tbl:res1}
\begin{center}
\begin{tabular}[t]{llccccccc} 
    Code & Symmetry   & $n_{\rm unique}$ & 
    $n_{\rm max}$ & $n_{\rm symm}$ & $n_{\rm T}$ &
$N_{\rm MC}$\tablenote{From Ref.~\onlinecite{Deem92a}.} & $N_{\rm BMC} $ & $N_{\rm PT}$
\\[.2em] \hline
ABW & P n a $2_1$     & 2 & 1 & 4 & 8 & 1 & 1 &  \\
ACO & I m $\bar{3}$ m & 1 & 1 & 96 & 16 & 1 & 1 &  \\
AEI & C 1 2/c 1       & 6 & 3 & 8 & 48 & 183 & 1 &  \\
AEL & I b m 2         & 6 & 3 & 8 & 40 & -- & 2 &  \\
AET & C m c $2_1$     & 10 & 5 & 8 & 72 & -- & 1 &  \\
AFG & P $6_3$/m m c   & 3 & 3 & 24 & 48 & 80 & 4 &  \\
AFI & P 6 c c         & 2 & 1 & 12 & 24 & 1 & 2 &  \\
AFO & C m c m         & 4 & 4 & 16 & 40 & -- & 2 &  \\
AFR & P m m n         & 4 & 4 & 8 & 32 & -- & 1 &  \\
AFS & P 3 c 1         & 12 & 3 & 6 & 56 & -- & -- & 1 \\
AFT & P $\bar{3}$ 1 c & 6 & 3 & 12 & 72 & 20 & 2 &  \\
AFX & P $\bar{3}$ 1 c & 4 & 2 & 12 & 48 & -- & 1 &  \\
AFY & P $\bar{3}$     & 4 & 2 & 6 & 16 & 20 & 4 &  \\
AHT & C m c m         & 2 & 2 & 16 & 24 & 24 & 1 &  \\
ANA & I a $\bar{3}$ d & 1 & 1 & 96 & 48 & 20 & 1 &  \\
APC & P b c a         & 4 & 2 & 8 & 32 & 20 & 1 &  \\
APD & P c a 2$_1$     & 8 & 2 & 4 & 32 & -- & 1 &  \\
AST & F 2 3           & 4 & 2 & 48 & 40 & -- & 1 &  \\
ATI & R $\bar{3}$     & 2 & 1 & 18 & 36 & -- & 1 &  \\
ATN & I $\bar{4}$     & 2 & 1 & 8 & 16 & 1 & 1 &  \\
ATS & C m c m         & 3 & 3 & 8 & 24 & -- & 1 &  \\
ATT & P $2_1$ $2_1$ 2 & 6 & 2 & 4 & 24 & 1 & 1 &  \\
ATV & A c m m         & 2 & 2 & 16 & 24 & 1 & 1 &  \\
AWW & P 4/n c c Z     & 4 & 2 & 16 & 48 & 20 & 5 &  \\
BEA & P $4_1$ 2 2     & 9 & 9 & 8 & 64 & -- & 5 &  \\
BEB & C 2/c           & 9 & 9 & 8 & 64 & -- & 4 &  \\
BIK & P 1             & 6 & 2 & 1 & 6 & 1 & 4 &  \\
BOG & I m m a         & 6 & 6 & 16 & 96 & 40 & 3 &  \\
BPH & P 3 2 1         & 6 & 3 & 6 & 28 & 116 & 6 &  \\
BRE & P $2_1$/m       & 4 & 4 & 4 & 16 & 1 & 1 &  \\
CAN & P $6_3$         & 2 & 1 & 6 & 12 & 1 & 1 &  \\
CAS & A m a 2         & 3 & 3 & 8 & 24 & 20 & 4 &  \\
CFI & I m m a         & 5 & 5 & 16 & 32 & -- & 2 &  \\
CHA & R $\bar{3}$ m R & 1 & 1 & 12 & 12 & 1 & 1 &  \\
CON & C 2/m           & 7 & 7 & 8 & 56 & -- & 1 &  \\
CZP & P $6_1$ 2 2     & 3 & 3 & 12 & 24 & -- & 1 &  \\
DAC & C 2/m           & 4 & 4 & 8 & 24 & 40 & 1 &  \\
DFO & P 6/m m m       & 6 & 6 & 24 & 132 & -- & 1 &  \\
DOH & P 6/m m m       & 4 & 4 & 24 & 34 & 80 & 2 &  \\
EAB & P $6_3$/m m c   & 2 & 2 & 24 & 36 & 1 & 1 &  \\
EDI & P $2_12_12$     & 3 & 2 & 4 & 10 & 1 & 1 &  \\
EMT & P $6_3$/m m c   & 4 & 4 & 24 & 96 & 20 & 1 &  \\
EPI & C 2/m           & 3 & 3 & 8 & 24 & 1 & 3 &  \\
ERI & P $6_3$/m m c   & 2 & 2 & 24 & 36 & 20 & 2 &  \\
EUO & C m m a & 10    & 10 & 16 & 112 & -- & -- & 1 \\
FAU & F d $\bar{3}$ Z & 2 & 1 & 96 & 192 & 1 & 1 &  \\
FER & I m m m         & 4 & 4 & 16 & 36 & 40 & 1 &  \\
GIS & P 1 1 2$_1$/a   & 4 & 1 & 4 & 16 & 2 & 1 &  \\
GME & P $6_3$/m m c   & 1 & 1 & 24 & 24 & 1 & 1 &  \\
GOO & P 1 $2_1$ 1     & 8 & 5 & 2 & 16 & -- & 4 &  \\
  \end{tabular}
\end{center}
\end{table}

\begin{table}[htbp]

\caption{Results for zeolite frameworks H to 
Y.\protect\tablenote{Legend as in Table \ref{tbl:res1}.}}
\label{tbl:res2}
\begin{center}
\begin{tabular}[t]{llccccccc} 
    Code & Symmetry   & $n_{\rm unique}$ & 
    $n_{\rm max}$ & $n_{\rm symm}$ & $n_{\rm T}$ & $N_{\rm 
MC}$\tablenote{From Ref.~\onlinecite{Deem92a}.} & 
$N_{\rm BMC}$ & $N_{\rm PT}$ \\[.2em] \hline
HEU & C 2/m           & 5 & 5 & 8 & 36 & 20 & 2 &  \\
IFR & C 2/m           & 4 & 4 & 8 & 32 & -- & 1 &  \\
ITE & C m c m         & 4 & 4 & 16 & 64 & -- & 1 &  \\
JBW & P m m a         & 2 & 2 & 4 & 6 & 1 & 3 &  \\
KFI & I m $\bar{3}$ m & 1 & 1 & 96 & 96 & 1 & 1 &  \\
LAU & C 2/m           & 3 & 3 & 8 & 24 & -- & 1 &  \\
LEV & R $\bar{3}$m    & 2 & 2 & 36 & 54 & 20 & 2 &  \\
LIO & P $\bar{6}$m2   & 4 & 4 & 12 & 36 & 20 & 1 &  \\
LOS & P $6_3$ m c     & 2 & 2 & 12 & 24 & -- & 1 &  \\
LTA & F m $\bar{3}$ c & 2 & 1 & 192 & 192 & 23 & 2& \\
LTL & P 6/m m m       & 2 & 2 & 24 & 36 & 1 & 1 &  \\
LTN & F d$\bar{3}$m   & 4 & 4 & 192 & 768 & -- & 1 &  \\
MAZ & P 6$_3$/m m c   & 2 & 2 & 24 & 36 & 1 & 1 &  \\
MEI & P 6$_3$/m & 4   & 4 & 12 & 34 & 80 & 5 &  \\
MEL & I $\bar{4}$ m 2 & 7 & 7 & 16 & 96 & -- & 6 &  \\
MEP & P m $\bar{3}$ n & 3 & 3 & 48 & 46 & -- & -- & 1 \\
MER & I m m m & 2 & 1 & 16 & 32 & 20 & 1 &  \\
MFI & P n m a & 12 & 12 & 8 & 96 & -- & -- & 1 \\
MFS & I m m 2         & 8 & 8 & 8 & 36 & -- & -- & 1 \\
MON & I $4_1$/a m d Z & 1 & 1 & 32 & 16 & 1 & 2 &  \\
MOR & C m c $2_1$     & 6 & 4 & 8 & 48 & 1 & 1 &  \\
MTN & F d $\bar{3}$m  & 3 & 3 & 192 & 136 & -- & 1 &  \\
MTT & P m m n         & 7 & 7 & 8 & 24 & -- & 1 &  \\
MTW & C 2/m           & 7 & 7 & 8 & 56 & -- & -- & 1 \\
MWW & P 6/m m m       & 8 & 8 & 24 & 72 & -- & -- & 3 \\
NAT & F d d 2         & 3 & 2 & 16 & 40 & 20 & 1 &  \\
NES & P 1 $2_1$/c 1   & 17 & 7 & 4 & 68 & -- & -- & 1 \\
NON & F m m m         & 5 & 5 & 32 & 88 & -- & -- & 1 \\
OFF & P $\bar{6}$ m 2 & 2 & 2 & 12 & 18 & 1 & 2 &  \\
PAU & I m$\bar{3}$ m  & 8 & 8 & 96 & 672 & -- & -- & 2 \\
PHI & P 1$2_1$/m 1    & 4 & 2 & 4 & 16 & 20 & 1 &  \\
RHO & I m$\bar{3}$m   & 1 & 1 & 96 & 48 & 1 & 1 &  \\
RTE & C 2/m           & 3 & 3 & 8 & 24 & 1 & 1 &  \\
RTH & C 2/m           & 4 & 4 & 8 & 32 & -- & 1 &  \\
RUT & P 1$2_1$/a 1    & 9 & 5 & 4 & 36 & -- & 1 &  \\
SAO & I $\bar{4}$ m 2 & 4 & 4 & 16 & 56 & -- & 1 &  \\
SAT & R $\bar{3}$ m   & 2 & 2 & 36 & 72 & -- & 1 &  \\
SBE & I 4/m m m & 4   & 4 & 32 & 128 & -- & 1 &  \\
SBS & P $\bar{3}$ 1 c & 8 & 4 & 12 & 96 & -- & -- & 1 \\
SBT & R $\bar{3}$ m   & 4 & 4 & 36 & 144 & -- & 1 &  \\
SGT & I $4_1$/a m d Z & 4 & 4 & 32 & 64 & 20 & 1 &  \\
SOD & P $\bar{4}$ 3 n & 2 & 1 & 24 & 12 & 1 & 2 &  \\
STI & C 1 2/m 1 & 5 & 4 & 8 & 36 & 40 & 4 &  \\
TER & C m c m & 8 & 8 & 16 & 80 & -- & 1 &  \\
THO & P n c n & 6 & 3 & 8 & 36 & -- & 3 &  \\
TON & C m c$2_1$ & 4 & 4 & 8 & 24 & -- & 1 &  \\
TSC & F m $\bar{3}$ m & 2 & 2 & 192 & 384 & -- & 1 &  \\
VET & P $\bar{4}$ & 5 & 5 & 4 & 17 & -- & 1 &  \\
VFI & P $6_3$/m c m & 2 & 2 & 24 & 36 & 1 & 1 &  \\
YUG & P 1 c 1 & 8 & 2 & 2 & 16 & 2 & 2 & \\  
  \end{tabular} 
\end{center}
\end{table}

\begin{table}[htbp]
\caption{Frameworks not attempted.}
\label{tbl:not}
\begin{center}
\begin{tabular}[t]{cccc} 
Open & With Templates & With Heavy Cations & Other \\ \hline
-CHI & CGF & LOV & WEI\\
-CLO & CGS & RSN &\\
-RON & DDR & VNI &\\
-WEN & ZON & VSV &\\ 
\end{tabular}
\end{center}
\end{table}

\normalsize

\end{document}